\documentclass[twocolumn, aps,amsfonts,asmsymb]{revtex4}

\begin{document}

\title{Cloning of orthogonal mixed states entails irreversibility}

\author{Micha{\l} Horodecki, Aditi Sen(De), and Ujjwal Sen} 
\affiliation{Institute of Theoretical 
Physics and Astrophysics, University of Gda{\' n}sk, 80-952 Gda{\' n}sk, Poland}

\begin{abstract}

Orthogonal pure states can be cloned as well as deleted. However if there is an initial disorder in the 
system, that is for orthogonal mixed states, one cannot perform deletion. And cloning, in such cases, necessarily 
produces an irreversibility, in the form of leakage of information into the environment.

\end{abstract}

\maketitle
 
Nonorthogonal states cannot be cloned \cite{WZ}. 
And orthogonal states can be cloned. However we will show that 
if there is an initial disorder in the system, one necessarily produces ``irreversibility'', even 
when cloning orthogonal states.

Consider two arbitrary mixed orthogonal states. For definiteness, we take them to be of rank two each. This restriction
does not change the generality of the statements that are made below. Suppose therefore that the two orthogonal
 mixed states are
\begin{equation}
\label{statesgulo}
\begin{array}{lcl}
\varrho_0 = p \left|0\right\rangle \left\langle 0 \right| + q \left| 2 \right\rangle \left\langle 2 \right|,  \\
\varrho_1 = r \left|1\right\rangle \left\langle 1 \right| + s \left| 3 \right\rangle \left\langle 3 \right|,
\end{array}
\end{equation}
where of course \(p\), \(q\), \(r\), \(s\) are nonnegative numbers such that \(p+q=1\), and \(r+s=1\), and
\(\left|0\right\rangle\), \(\left|1\right\rangle\), \(\left|2\right\rangle\), \(\left|3\right\rangle\) are a set of 
mutually orthonormal
states.

Consider now the  task of  cloning the two states, \(\varrho_0\) and \(\varrho_1\), if any one is given.
So we want to create \(\varrho_i \otimes \varrho_i\) from \(\varrho_i\) coupled with a blank state (that is, a state
 that has no information about
\(i\), \(i = 0,1\)).
There is a trivial way to do that. One simply makes a measurement onto the (rank-two) projection operators
\[
\begin{array}{lcl}
P_0 = \left| 0 \right\rangle \left\langle 0 \right| + \left| 2 \right\rangle \left\langle 2 \right|,   \\
P_1 = \left| 1 \right\rangle \left\langle 1 \right| + \left| 3 \right\rangle \left\langle 3 \right|.
\end{array}
\]
If \(P_0\) clicks, then the conclusion is that the given state was \(\varrho_0\). Otherwise, the state was \(\varrho_1\).
After finding out what the state is, one can just prepare the required extra copy of the state that is indicated by the
 measurement.
Note that one may deliberately perform the finer measurement onto the orthonormal basis consisting of the states
\(\left|0\right\rangle\), \(\left|1\right\rangle\), \(\left|2\right\rangle\), \(\left|3\right\rangle\). That would
 destroy the information about the mixing probabilities (\(p\), \(q\), and \(r\), \(s\)) in the states
\(\varrho_0\) and \(\varrho_1\). But we assume that these mixing probabilities are known. So if, for example,
\(\left|2\right\rangle\) clicks in such a finer measurement, the conclusion is that the given state was 
\(\varrho_0\). One can then prepare two copies of \(\varrho_0\).

However, for the case when \(q = s = 0\), that is when \(\varrho_0\) and \(\varrho_1\) are pure, and are
respectively   \(\left|0\right\rangle \left\langle 0 \right|\) and    \(\left|1\right\rangle \left\langle 1 \right|\), 
there is
another way to clone. It is by using a gate that takes
\begin{equation}
\label{pure_clone}
\begin{array}{lcl}
\left|0\right\rangle_S  \left|0\right\rangle_B  
																	\rightarrow \left|0\right\rangle_S \left|0\right\rangle_B, \\
\left|1\right\rangle_S \left|0\right\rangle_B 
											    					\rightarrow \left|1\right\rangle_S \left|1\right\rangle_B.
\end{array}
\end{equation}
This could for example be the CNOT gate, which additionally takes
\[
\begin{array}{lcl}
\left|0\right\rangle  \left|1\right\rangle  \rightarrow \left|0\right\rangle \left|1\right\rangle, \\
\left|1\right\rangle \left|1\right\rangle \rightarrow \left|1\right\rangle \left|0\right\rangle.
\end{array}
\]
To see the cloning of the states, consider the states marked by \(S\)  in eq. (\ref{pure_clone}) as the 
original copy, and those marked by  \(B\)  as the blank copy. One then obtains the 
two copies of the original on the right-hand-side of eq. (\ref{pure_clone}).

Note an important difference in the two ways of cloning. In the first method, a measurement step is involved.
This measurement, results in making the process ``open''. After the whole process of cloning has been completed,
there is information about the system 
(precisely, the result of the measurement) left in the environment. Let us call this method of cloning orthogonal 
states as ``open cloning''.

Now contrast this method of cloning with the second method of cloning orthogonal \emph{pure} states, by using 
CNOT-type gates. In this case, after the cloning process has been completed, there is no information
about the system that is left in the environment. Let us call this method of cloning orthogonal \emph{pure} states
as ``closed cloning''. 

Open cloning is thus an irreversible process. It leads to production of some ``garbage'' in the 
environment. Closed 
cloning, on the other hand, is a ``clean'' process. There is no left out garbage in this case. We will show 
that cloning of orthogonal mixed states will always produce  garbage. Such states cannot be cloned by 
closed cloning.

Consider the state \(\varrho_i\) (\(i = 0,1\)). We want to produce two copies of 
\(\varrho_i\). That is, we want to produce \(\varrho_i \otimes \varrho_i\). At the input, we take a blank 
copy \(\varrho_b\) (along with \(\varrho_i\)). 
Therefore our input is \(\varrho_i \otimes \varrho_b\), \(i = 0,1\).  And we want \(\varrho_i \otimes \varrho_i\)
at the output. If we do not want the information about \(i\)  to get leaked into the environment, this 
evolution must be done unitarily. Clearly, this cannot be done. To see this, just note that unitary evolution preserves 
the spectrum. But \(\varrho_i \otimes \varrho_b\) cannot have the same spectrum as \(\varrho_i \otimes \varrho_i\)
for both \(i = 0\) \emph{and} \(1\). Therefore cloning of two orthogonal mixed states is not possible in a closed 
system. To implement this cloning, one is bound to do a measurement, so that the cloning is an open cloning.

Let us now consider the deleting process. 
In deleting \cite{Pati}, one requires to have \(\varrho_i \otimes \varrho_b\) at the output, when 
\(\varrho_i \otimes \varrho_i\) is fed at the input. And we know that deleting must necessarily 
be considered in a closed system (see \cite{Jozsa,amader-cloning} in this regard). So the same reasoning
as above, renders 
such deleting impossible by a unitary evolution.  Thus we have that deleting of orthogonal mixed 
states is not possible.

In the case of orthogonal pure states (\(\left|0\right\rangle\) and \(\left|1\right\rangle\)), one can clone (as in 
eq. (\ref{pure_clone})), and by the inverse operation, one can delete. To see this, note that the inverse operation 
takes  
\[
\begin{array}{lcl}
\left|0\right\rangle  \left|0\right\rangle  \rightarrow \left|0\right\rangle \left|0\right\rangle, \\
\left|1\right\rangle \left|1\right\rangle \rightarrow \left|1\right\rangle \left|0\right\rangle.
\end{array}
\]
However in the case of orthogonal mixed states, considered in this paper, one cannot delete, while cloning is 
possible if one considers an open system. Therefore for orthogonal \emph{mixed} states, cloning and deleting are 
not inverse processes of each other.

\end{document}